\documentclass[%
 reprint,
 amsmath,amssymb,
 aps,
]{revtex4-2}

\usepackage{multirow}
\usepackage{romannum}
\usepackage{color}
\usepackage{graphicx}
\usepackage{dcolumn}
\usepackage{bm}
\usepackage{siunitx}
\usepackage[normalem]{ulem}

\begin{document}

\preprint{APS/123-QED}

\title{
Tailored elastic surface to body wave Umklapp conversion}

\author{Gregory~J.~Chaplain$^{1}$,  
 Jacopo M. De Ponti$^{2,3}$, Andrea Colombi$^4$, \\
 , Rafael Fuentes-Dominguez$^{5}$, Paul Dryburg$^{5}$, Don Pieris$^{5}$, Richard J. Smith$^{5}$, Adam Clare$^{6}$, Matt Clark$^{5}$ and Richard~V.~Craster$^{1,7}$
}
\affiliation{$^1$ Department of Mathematics, Imperial College London, London SW7 2AZ, UK}
\affiliation{$^2$ Dept. of Civil and Environmental Engineering, Politecnico di Milano, Piazza Leonardo da Vinci, 32, 20133 Milano, Italy}
\affiliation{$^3$ Dept. of Mechanical Engineering, Politecnico di Milano, Via Giuseppe La Masa, 1, 20156 Milano, Italy}
\affiliation{$^4$ Dept. of Civil, Environmental and Geomatic Engineering ,ETH, Stefano-Franscini-Platz 5, 8093 Z\"urich, Switzerland}
\affiliation{$^5$
Optics and Photonics, Faculty of Engineering, University of Nottingham, Nottingham, NG7 2RD, UK}
\affiliation{$^6$
Advanced Component Engineering Laboratory (ACEL), Faculty of Engineering, University of Nottingham, NG7 2RD, Nottingham, UK}
\affiliation{$^7$ Department of Mechanical Engineering, Imperial College London, London SW7 2AZ, UK}

\maketitle
\textbf{
Elastic waves guided along surfaces dominate applications in geophysics, ultrasonic inspection, mechanical vibration, and surface acoustic wave devices; precise manipulation of surface Rayleigh waves and their coupling with polarized body waves presents a challenge  that offers to unlock the flexibility in wave transport required for efficient energy harvesting and vibration mitigation devices.  
We design elastic metasurfaces, consisting of a graded array of rod resonators attached to  an elastic substrate that, together with  critical insight from Umklapp scattering in phonon-electron systems, allow us to leverage the transfer of crystal momentum; we mode-convert Rayleigh surface waves into bulk waves that form tunable beams. 
 Experiments, theory and simulation verify that these  tailored Umklapp mechanisms play a key role in coupling surface Rayleigh waves to reversed bulk shear and compressional waves independently, thereby creating passive self-phased arrays allowing for tunable redirection and wave focusing within the bulk medium.
  } 

The Umklapp, or flip-over, process, first hypothesised by Peierls \cite{Peierls_Thesis} is conventionally concerned with describing phonon-phonon scattering to explain thermal conductivity at high temperatures, and has a rich history in the quantum theory of thermal transport \cite{Hoddeson87development,taylor2002quantum}. We take concepts based around the Umklapp process into an area of wave physics, where it is not traditionally used, and use it to provide the insight required to design a class of elastic metasurfaces; deep elastic substrates support surface Rayleigh waves that propagate along the surface, often over large distances, which are an essential component of, for instance, surface acoustic wave microfluidic devices \cite{yeo2014}, acoustic microscopy \cite{briggs92a}, at small-scales and of seismic wave and groundborne vibration propagation at the geophysical scale \cite{dahlen98a}. An isotropic, and homogeneous, elastic medium supports two types of bulk waves: compressional, P,  and shear, SV and SH, waves polarised vertically and horizontally that propagate with different wavespeeds $c_p$ and $c_s$ with $c_p>c_s$ \cite{graff75a} with the Rayleigh wavespeed $c_r$ slower than both. 

We seek to combine the Umklapp process with recently emerging ideas in graded metamaterial arrays and so-called rainbow trapping devices. To do so we take advantage of ideas that emerged in optics around slow-light devices and optical ``rainbow" trapping \cite{Tsakmakidis2007} that have in turn motivated tailored designs of graded Helmholtz resonator arrays in acoustics \cite{Zhu2013}, and their analogues in water waves \cite{bennetts18a}, to slow the array guided waves and trap the waves at different spatial positions, with application to broadband sound absorbers \cite{Romero-Garcia2013,Jimenez2017}. 
  
Neither acoustic or electromagnetic waves have the additional complications of elasticity, that is, having both shear and compressional wavespeeds, mode coupling at interfaces and Rayleigh surface waves; this presents an opportunity as the additional degree of freedom in elasticity can be exploited. 
Elastic graded resonator arrays use rods or beams, whose length determines the resonance frequency, and grading creates a metawedge \cite{colombi16a}. Trapping, and slow-wave, phenomena occur  but now with the additional physics of mode conversion from the Rayleigh wave into a forward-propagating shear wave in the bulk as confirmed experimentally \cite{Colombi2017}. The trapping phenomena has potential for energy harvesting \cite{celli1};  the metawedge provides spatial segregation by frequency thereby amplifying elastic energy at specific resonators that can be coupled to piezoelectric patches \cite{DePonti2019,Alan2019}.
\begin{figure*}[t]
    \centering
    \includegraphics[width = 0.95\textwidth]{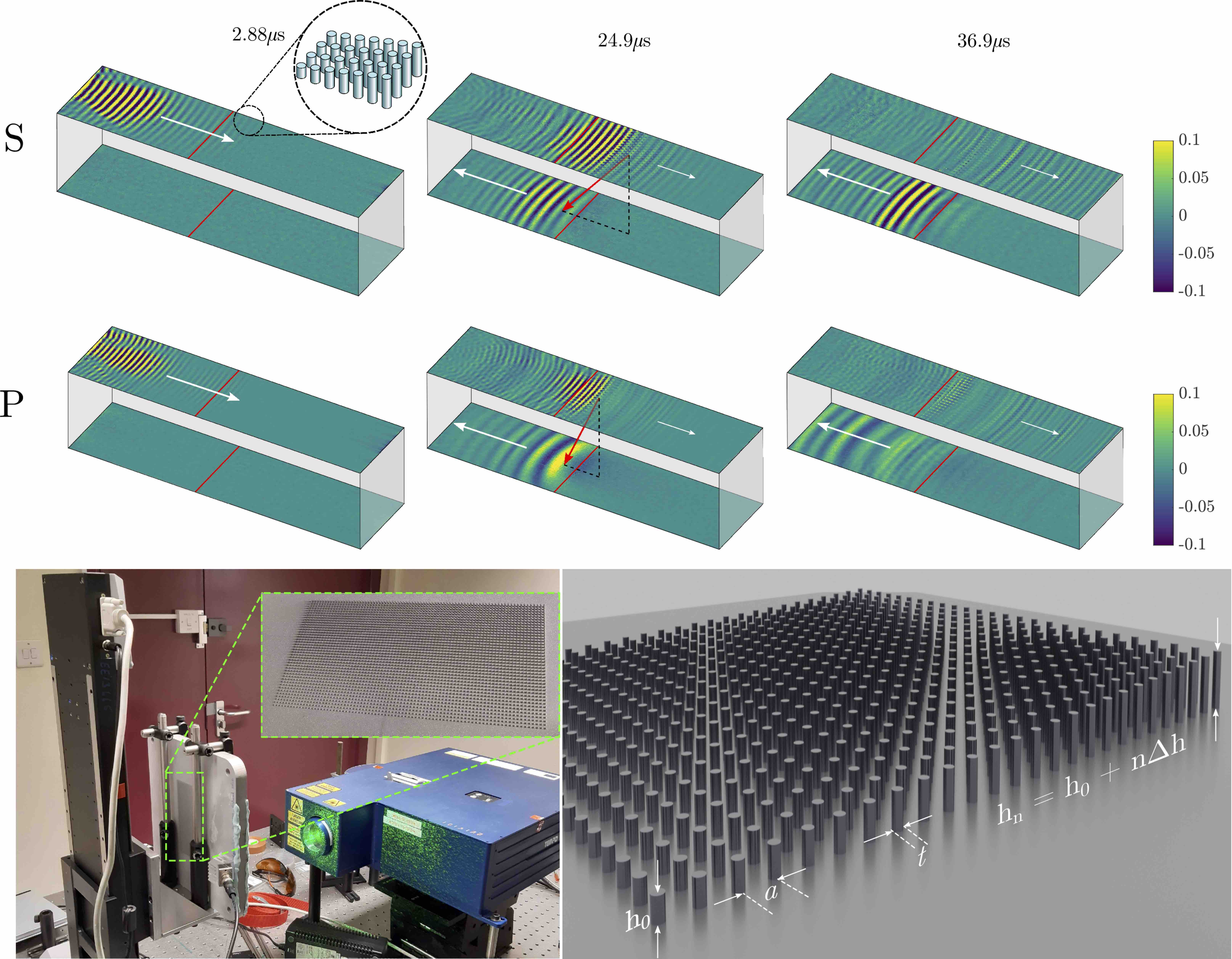}
    \caption{{Experimental results:} snapshots in time of temporal-spatially filtered scans along top and bottom surfaces observing S conversion (top row) and P conversion (bottom row), filtered between $1.1-1.2$MHz and $1.45-1.55$MHz respectively, normalised to the maximum of displacement of the top surface. Solid red lines show where graded array begins, with increasing rod height in the direction of wave propagation on the top surface. The reversed conversion is strikingly clear; the forward propagating surface wave on the top surface is reverse converted into the bulk and is seen to excite reversed propagating surface waves on the bottom surface. The measured angles of reversed conversion, $-131.8^{\circ}$ and $-106.9^{\circ}$ for S and P respectively, match the predicted angles from Fig.~\ref{fig:SP}. The experimental setup is shown, as is a schematic of array geometry detailing the rod diameter, $t$, periodicity $a$ and grading through the changes in height of the $n$th rod, $h_n$, as $h_n=h_0+\Delta h$. Fabrication details are given in the methods section, with the array parameters given in Table~S1 of the supplemental material, \textit{with videos of the experimental results available in the online supplemental material}.}
    \label{fig:3D_top_bottom}
\end{figure*}

The physics of a graded structured array can also be interpreted in terms of phase changes, induced by changes in the array elements, as the wave transits the array. In this vision, a graded array acts as a self-phased array and we can induce backward directed radiation from an array. In the context of a model for flexural waves on thin elastic plates, a simple scalar model, a graded line array  has been shown \cite{Chaplain2019Flat} to create focusing and flat-lensing effects, that emulate negative refraction on a line and Pendry-Veselago flat lenses. Although motivational, the thin plate model also lacks the multiple wavespeeds of the deep elastic substrate, and the array guided wave only exists below the sound-line, i.e. in the first Brillouin zone. 

\begin{figure*}
\includegraphics[width = 0.85\textwidth]{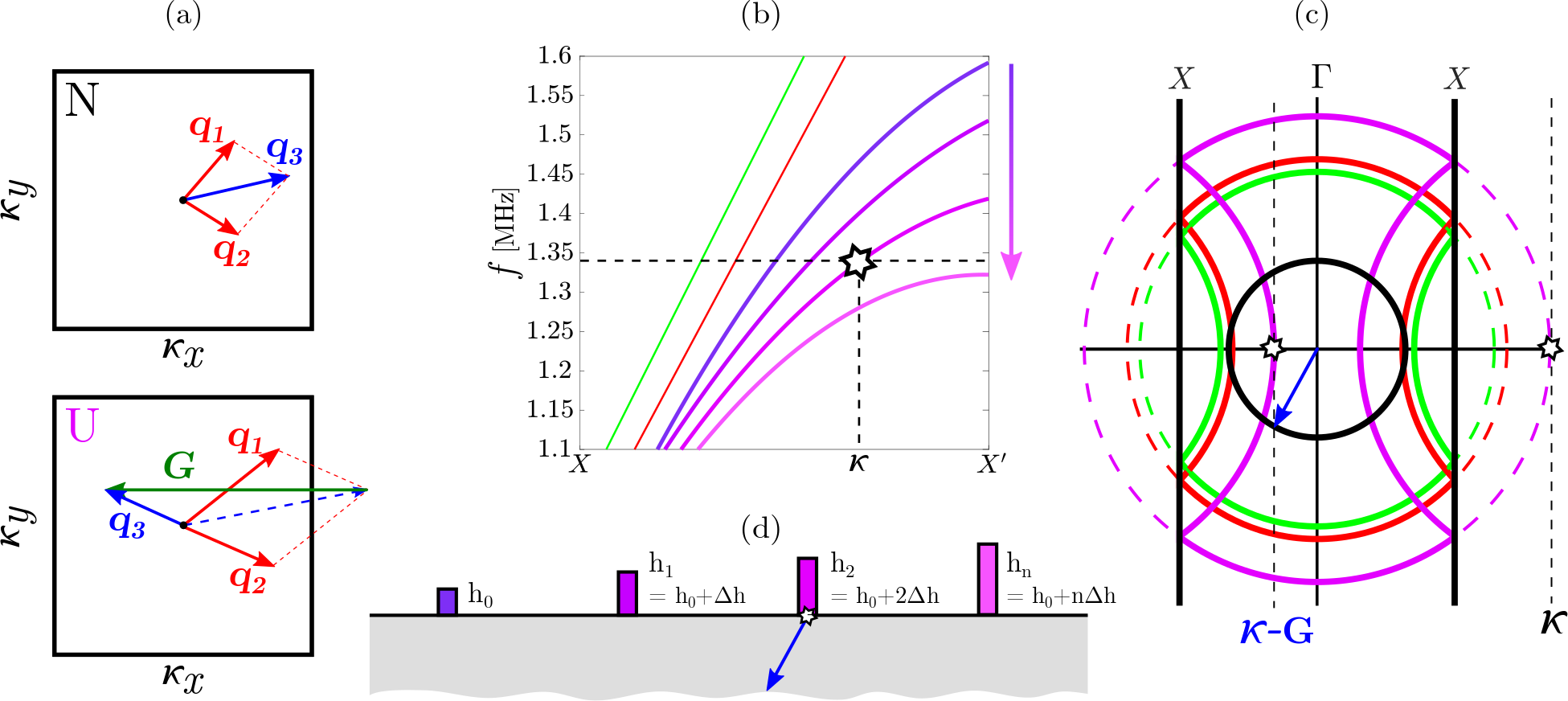}
\caption{(a) Wave vector scattering in the first Brillouin zone showing conventional definitions of an N-process (top) and a U-process (bottom); the 'flip-over' description is only strictly true for collinear scattering \cite{Maznev2014}. (b) Longitudinal dispersion curves within \textit{second} Brillouin zone; $X$ marks the edge of the first Brillouin zone, whilst $X'$ marks the edge of the second Brillouin zone. Each dispersion curve represents the second longitudinal mode for a perfectly periodic array of rods, with the fixed heights shown in (d), and with R(S) sound lines shown in red(green). (c) Isofrequency contours of rod where last longitudinal mode supported, after which an effective bandgap is reached, when exciting at a frequency of $1.340 \si{\mega\hertz}$. At this position U-processes are preferential to a change in rod motion and, resulting in an effective reversed wavenumber $\boldsymbol{\kappa} - \boldsymbol{G}$, capable of mode converting into a P wave (black isofrequency contour, shown in full dispersion curves in Supp. Mat Fig.~S1). (d) Graded array schematic, colours corresponding to the curves in (b).}
\label{fig:UmklappDispersion}
\end{figure*}

\begin{figure*}[t!]
\includegraphics[width = \textwidth]{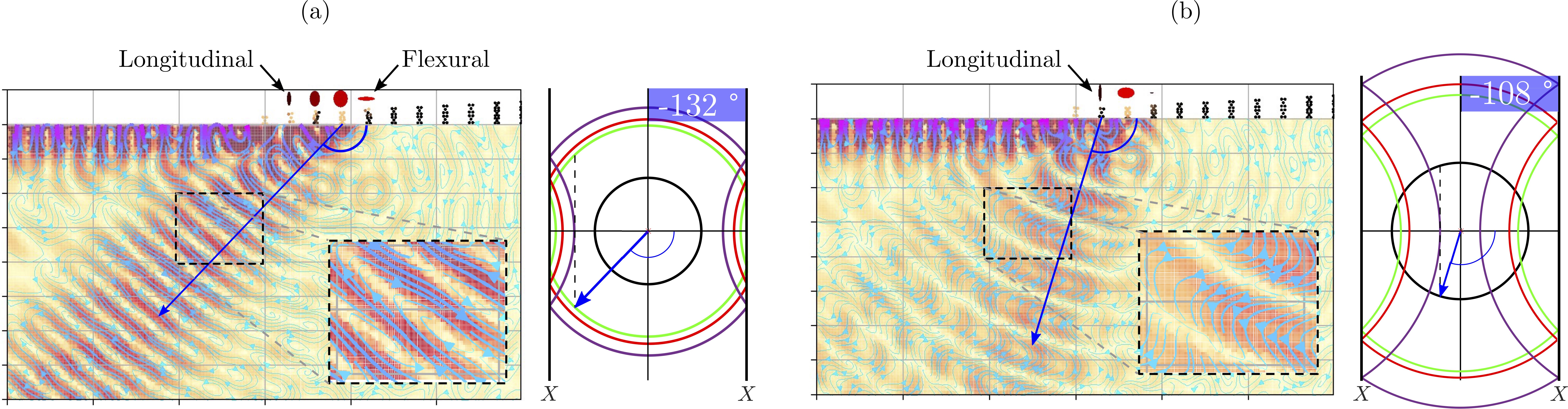}
\caption{Reversed conversion for the array detailed in Table~S1. (a) S conversion for excitation at $f = 1.2 \si{\mega\hertz}$ (b) P conversion for excitation at $f = 1.45 \si{\mega\hertz}$. In each case at the position of conversion is determined by the change from longitudinal to flexural motion of neighbouring rods. The ratio of the relative magnitude of longitudinal to flexural motion is shown by the ellipses above the rods. The angle of conversion, computed directly from the simulated field transformed in the wavenumber space, matches the predictions from the isofrequency curves of the last rods supporting the longitudinal mode, similarly to Fig.~\ref{fig:UmklappDispersion}(c). The difference between the converted wave types (S and P) can be seen clearly from the streamline analysis. Wavenumber analysis on the simulated data also confirmed the two different velocities approx $3000 \si{\meter\second}^{-1}$ for S and $6000\si{\meter\second}^{-1}$ for P.}
\label{fig:SP}
\end{figure*}

In this Letter, by taking an elastic half-space, patterned by a graded resonant array of rods (Fig.~\ref{fig:3D_top_bottom}), we show that, quite remarkably, one can obtain mode conversion from surface waves to compressional, P, waves, and not just to shear, S, waves, in the bulk. Furthermore, this can be directed backward, and used to create focusing, and these features are due to Umklapp scattering from outside the first Brillouin zone, i.e. from above the sound-line; this is counter-intuitive in that the waves above the light-line are traditionally ignored when considering spoof surface plasmons \cite{Pendry2004} which are the closest electromagnetic analogue to the Rayleigh wave. These striking results are confirmed experimentally in Fig.~\ref{fig:3D_top_bottom} from the predictions made by the theory, see Fig.~\ref{fig:UmklappDispersion}, and by simulation (Fig.~\ref{fig:SP}).
The elastic wave system, having  distinct wave types with  different wavespeeds \cite{achenbach2012wave}, is imbued with richer physics than acoustic/electromagnetic systems and this yields a greater degree of flexibility and the opportunity for novel effects; in elasticity we have two distinct sound-lines, also as the Rayleigh surface wave is, unlike spoof surface plasmons, not induced by periodic geometrical changes it is not confined to lie below the sound-line nor is it confined to be within the first Brillouin zone.


When operating outside the first Brillouin zone concepts of crystal momentum become important \cite{Maznev2014}. The conventional definition of an Umklapp process, or U-process, is elucidated in Fig.~\ref{fig:UmklappDispersion}(a), whereby the resultant wavevector of a scattering process within a periodic crystal lies outside the first Brillouin zone; promotion of a backward propagating phonon results through crystal momentum transfer since the wave vectors, $\boldsymbol{q}$, are defined modulo $\boldsymbol{\mathbf{G}}$ i.e. up to a reciprocal lattice vector. The textbook distinction between normal (N-processes, Fig.~\ref{fig:UmklappDispersion}(a)) and U-processes is then given by
\begin{equation}
    \boldsymbol{q}_{1} + \boldsymbol{q}_{2} - \boldsymbol{q}_{3} = \begin{cases}
    \boldsymbol{0} & \text{N-process}, \\
    \boldsymbol{G} & \text{U-process}.
    \end{cases}
    \label{eq:umklapp}
\end{equation}
where the $\boldsymbol{q}_{i}$ are the wave vectors in Fig.~\ref{fig:UmklappDispersion}(a). We stress that these mechanisms do not violate momentum conservation \cite{Peierls_Thesis}; unlike in conventional graded structures we are not considering only the true momenta of interfering waves (phonons) within a crystal, but taking advantage of the momentum of the system as a whole. Nuances of this simplistic definition arise due to the apparent interchangeability of N- and U-processes through altering the primitive cell, and the associated (quasi)momentum conservation \cite{peierls1996quantum,Maznev2014,ding2018umklapp}. The distinction is achieved throughout by analysing the dispersion curves of the locally periodic elements, along with simplified isofrequency contours (Supplemental Fig. S1(b)).  Recently the importance of the Umklapp process has been solidified for electron-electron scattering through experimental verification, highlighting the fundamental role it plays in electrical resistance in pure metals \cite{Wallbank2019}, along with the utility to probe electronic structures \cite{rader2004elastic}. Further to this, new breeds of entirely flat lens devices have been theorised which incorporate Umklapp scattering processes for surface waves on dielectric substrates \cite{Chaplain2019c} by virtue of abrupt, not adiabatic, gradings.


Undeterred by its prevalence in quantum, discrete systems we demonstrate the efficacy of U-processes which can preside over purely passive, classical, continuous elastic devices. The extent of this mechanism is often not considered in the wide range of structures based on (locally) periodic material systems, and indeed neglected in elasticity theories \cite{gurevich1990quasimomentum}. Incorporating this effect is achieved by utilising the existence of surface waves outwith the periodic structure, and marrying the transition of such a wave to the excited wave within a graded structuring. By doing so, striking reversed conversion into S and P waves can be achieved, and controlled, as predicted in Fig.~\ref{fig:SP} and experimentally verified in Fig.~\ref{fig:3D_top_bottom}. We present here the design methodology for structures capable of the conversion of Rayleigh waves directly into both S and P waves independently, by utilising the dispersion curves and isofrequency contours of perfectly periodic arrays of rods.

Shown in Fig.~\ref{fig:UmklappDispersion}(b), are the dispersion curves for the longitudinal motion of rods with fixed heights within the \textit{second} Brillouin zone. These curves are obtained through an averaging process of the fully polarised dispersion curves (Supplemental Fig.~S1), since incident R-waves excite a superposition of longitudinal and flexural rod motion \cite{Colquitt2017}. The behaviour of the adiabatically graded array, as shown in Fig.~\ref{fig:UmklappDispersion}(d) is predicted from these locally periodic curves. For a given frequency, longitudinal motion is preferentially supported by a number of rods; at some rod height however, there is a transition in preferred rod motion from longitudinal to flexural (or vice versa). This is seen as an effective band gap in Fig.~\ref{fig:UmklappDispersion}(b) for the fourth rod, where U-processes dominate;  the transfer of crystal momentum results in an effective reversed wavenumber, $\boldsymbol{\kappa} - \boldsymbol{G}$. Depending on the operational frequency, this can lie within the isocircle of the free S or P body waves; reversed conversion is achieved by conservation of the tangential component of the wavevector. Figure~\ref{fig:UmklappDispersion}(c) shows the prediction of this angle by inspection of the last supported longitudinal mode (corresponding to the third resonator). The projected dispersion curves show the resultant wave vector inside the first Brillouin zone, as a result of transfer of crystal momentum, by the white star corresponding to that in (a). The wave can hybridise with a reversed P wave by the phase matching with the isofrequency circle of the P wave, shown in black. A similar analysis can be carried out for conversion into S waves, by operating at a lower frequency. In this way, either longitudinal or flexural rod motion can excite either S or P reversed waves by the inspection of the transition from one dominant rod motion to another.

When operating at higher frequencies, outside the first Brillouin zone, U-processes occur regardless of any grading; energy is continually shed along the array (Supplemental Fig. S1(c)), resembling an elastic leaky wave antenna \cite{Balanis2008}. This can be manufactured to take place at a desired position by either adiabatic or abrupt gradings \cite{Chaplain2019c}. A similar effect is observed when exciting along the opposite array direction; the grading experienced is then from high to low rods. In this case, since the effective bandgap is not a true bandgap, propagation occurs through the array with Umklapp processes taking place all along the array (supplementary Fig.~S5).

Within this simple metawedge design paradigm lies many degrees of freedom: the position, angle and wave-type of reversed conversion, all allow a tailored conversion which simply relies upon a change in primary mode behaviour between neighbouring rods, within a higher BZ. Demonstrated in Fig.~\ref{fig:SP}, modelled using SPECFEM \cite{peter2011forward}, is the reversed conversion into S and P modes via the transition from longitudinal to flexural rod motion for an array with parameters detailed in Table~S1. The ellipses above the rods convey the ratio of the two dominant rod behaviours via their semi minor and major axis. The angle of conversion comes from the conserving the tangential component of the wavevector, determined via the isofrequency contours of the rod prior to the change in rod motion, showing S conversion for $1.2 \si{\mega\hertz}$ and P conversion for $1.45 \si{\mega\hertz}$. 

Simulation results are confirmed experimentally in a $1.8 \si{\centi\meter}$ thick slab of aluminum patterned using 3D printing, with a graded array of aluminum microrods on the surface (Fig.~\ref{fig:3D_top_bottom}, Supplemental Fig.~S5). The sample is mounted on a moving stage; a laser adaptive photorefractive interferometer scans the surface of the block and records the displacement field in the out of plane direction $u_z$. An ultrasonic transducer generates, via phase matching, purely Rayleigh waves at the surface of the aluminium block (see Methods for details on the experimental set-up). To enhance the visualisation of the conversion spatial and frequency filters have been applied. Time series have been bandpassed between $1.1 - 1.2 \si{\mega\hertz}$ and $1.45 - 1.55 \si{\mega\hertz}$ for S and P conversion respectively. The wavefield scans have been filtered selecting wavevectors pointing towards (away) from the resonators for the top (bottom) surface. This procedure mainly remove echoes from the boundaries as well as leakage from the transducer. The wavelength difference between input Rayleigh and converted S-waves suggests that the wavefront is mainly reflected upward in the bulk according to Snell's law only partly converting into backward travelling Rayleigh waves along the bottom surface. In the P conversion case, the effect is exacerbated and differences in wavelength, velocity and propagation direction between top and bottom signal are striking clear.

To conclude, we have shown that crystal momentum transfer via Umklapp scattering is of paramount importance to furthering the modalities of many metamaterial devices. Leveraging this decades old phenomenon with modern, advanced structured materials permits these remarkable devices to harness further powers of wave manipulation. For elasticity we have experimentally verified a tailored surface wave to body wave reversed conversion effect, which allows the distinct bulk wave-types to be separated at will. The incorporation of these mechanisms motivates hybrid effects with self-phased systems, with potential to spark new paradigms of control across all wave regimes.

\textbf{\textit{Methods:}} The sample was fabricated by selective laser melting (SLM) and measured by a rough-surface capable optical detector (Bossanova Tempo). The ultrasound was generated by using a Panametrics Accuscan A402S 1 MHz plane wave transducer with a 65$^{\circ}$ polymer wedge to couple the longitudinal wave of the transducer into a Rayleigh wave on the sample. The transducer and the wedge were glued to the sample using phenyl salicylate which provided good coupling and long term stability. The transducer was driven by a Ritec RPR-4000 programmable pulser using a 3-cycle sinusoidal burst at 1 and 1.5~MHz for S and P-conversion, respectively, with an amplitude of 300~V peak-to-peak and repetition rate of 500~Hz. At this repetition rate, all echoes from the previous pulses were able to die out before the next measurement. The signal was captured using a digital oscilloscope (Agilent) with 125~MSa/s and averaged 512 times before transfer to a desktop computer. The sample was mounted on scanning stages with a movement range of 100$\times$ 30 mm with a step-size of 0.25~mm.

The propagation of Rayleigh, P and S waves in a 2D halfspace patterned with surface resonators 
is modeled by solving the P-SV elastic wave equation. 
The 2D time domain simulations are carried out using SPECFEM2D, a code that solves the elastic wave equation using finite difference in time and the spectral element method in space. The parallelization is implemented through domain decomposition with MPI. The mesh is made of quadrilateral elements and it is generated using the commercial software Trelis. The top surface is traction free to support the propagation of surface waves, the other boundaries are supplied with perfectly matched layer conditions to mimic unbounded wave propagation and to avoid undesired reflections. 
Simulations are then run on a parallel cluster on 16 CPUs. 2D and 3D plots have been generated using Python library Maplotlib and Matlab. 

\textbf{\textit{Acknowledgements:}} The Authors would like to thank  Richard S\'{e}los for printing the sample. A.C. was supported by the Ambizione Fellowship PZ00P2-174009. The support of the UK EPSRC through grants EP/K021877/1 and EP/T002654/1 is acknowledged. 



%

\clearpage
\section{Tailored elastic surface to body wave Umklapp conversion: Supplementary Material}
\setcounter{figure}{0}
\renewcommand{\thefigure}{S\arabic{figure}} 
\renewcommand{\thetable}{S\arabic{table}} 

\begin{figure*}[bt!]
	\centering
	\includegraphics[width = 0.99\textwidth]{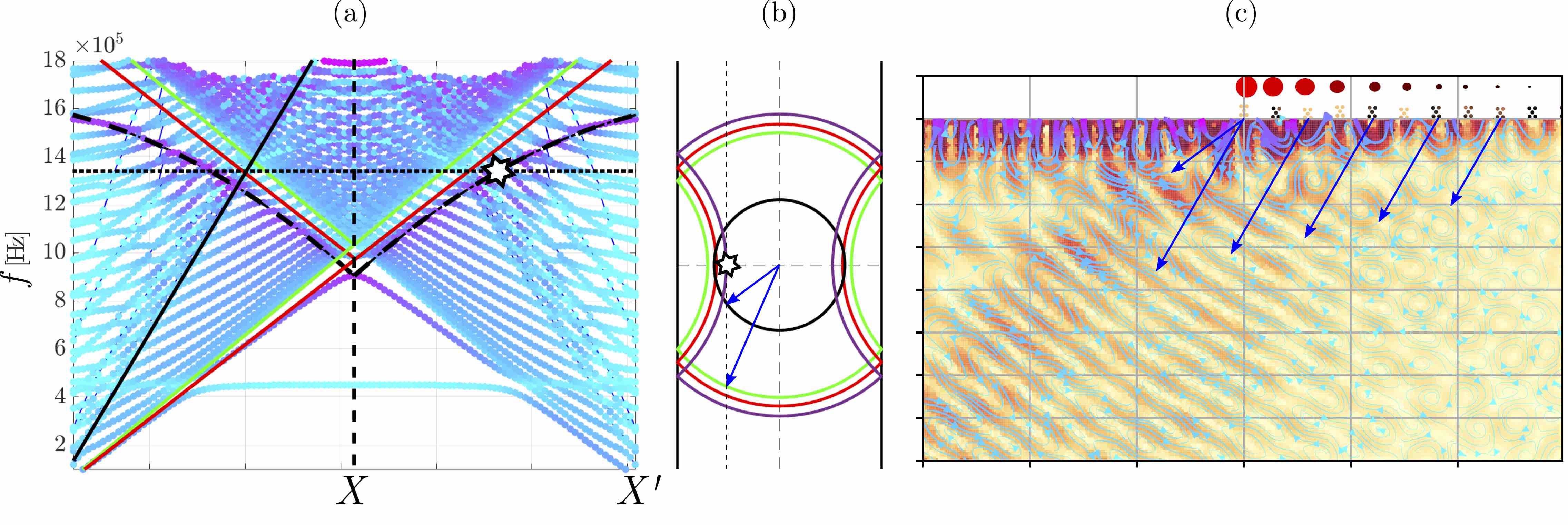}
	\caption{(a) A portion of the extended dispersion curves for an array of rods with parameters as in Table S1. The Rayleigh, shear and compressional sound lines are shown in red, green and black respectively. The relative dominance of longitudinal to flexural rod motion is shown by the colormap with purple/blue being predominantly longitudinal/flexural. Dashed black lines show the interpolated second longitudinal mode, with excitation frequency of $1.34 \si{\mega\hertz}$ used in Fig. 2 as the dotted horizontal line. Band folding occurs so that the IBZ is shown up to the dashed vertical black line. The excited mode wavevector is highlighted by the white star that, after the transfer of crystal momentum is reversed and the scattered wavevector is shown in the corresponding isofrequency contours in (b). Panel (c) shows scattering by an ungraded array consisting of perfectly periodic array of resonators, as defined in Table S1, with $\Delta h = 0$, excited by an incoming Rayleigh wave of frequency $1.34 \si{\mega\hertz}$. Umklapp scattering takes place without grading, but is not confined to a beam since no effective band gap reached; this is illustrated by the eccentricity of the ellipses above the rods remaining unchanged. A mixture of shear (S) and compressional (P) waves is produced 
		as predicted from the isofrequency curves in (b).}
	\label{fig:FullDisp}
\end{figure*}

To further elucidate the insight underlying how the reversed conversion effect is achieved, we present a detailed analysis of the full dispersion curves along with carefully chosen additional examples of the reversed conversion effect; experimental details are also provided. As an additional independent validation, scattering simulations were performed using the finite element software Abaqus \cite{abaqus2011abaqus} (not presented), as opposed to SPECFEM used within the main article.

We emphasise that Eq. (1), presented in the main text, does not violate momentum conservation \cite{Peierls_Thesis}; unlike for conventional graded structures we are not considering only the true momenta of interfering waves (phonons) within a crystal, but taking advantage of the momentum of the system as a whole. 
To utilise Umklapp U-processes for surface waves interacting with a graded structure with constant periodicity between resonant elements, it is paramount to obtain the dispersion curves for a perfectly periodic, infinite array of rods of fixed height; the adiabatic grading of the array allows the dispersion curves at each grading parameter to be used to infer the behaviour of the entire array; this assumption is now commonplace in the design of graded/chirped structures \cite{colombi16a,Romero-Garcia2013}. 

\textbf{\textit{Dispersion curves:}} In Fig.~\ref{fig:FullDisp} we show a portion of the extended dispersion curves, up to the edge of the second Brillouin zone, marked $X'$, for a perfectly periodic medium of resonant rods atop an elastic half space at fixed height $h = 0.5 \si{\milli\meter}$ where the diameter of the rods and periodicity of the array are given in Table S1. Using COMSOL multiphysics finite element software these dispersion curves, along with the relative rod motion (either flexural or longitudinal), are calculated. The rod's dominant behaviour is important in terms of how the surface wave hybridises into a bulk wave, and is shown through the colormap of the dispersion curves, with purple denoting longitudinal (axial) motion and blue displaying flexural motion. Conventionally when considering surface waves in periodic graded systems, only the lowest two modes below the Rayleigh line are considered \cite{colombi16a,Colombi2017}; these modes are clearly seen in Fig.~\ref{fig:FullDisp}(a) as they are strongly localised to the surface and therefore are clearly identifiable. For higher modes, whose frequencies are outside the first BZ, the dispersion curves for surface waves are more difficult to ascertain, due to the large number of spurious solutions \cite{Chaplain2019} found by the eigenvalue solver; these arise from the finite depth of the simulated region and correspond to propagating modes or modes created by the finite layer. The relative degree of longitudinal/flexural motion for higher modes is not as clear as for the lowest modes. To extract the excited, localised surface modes presented in Fig. 2(b), we use a weighted averaging technique to interpolate, and hence separate, both motions independently. We remove curves with the opposite behaviour below a set tolerance on the vertical and horizontal displacements of the rods; the ellipses above the rods in Fig. 3 represent this ratio. At each wavevector, the remaining modes are then averaged with a weight normalised to the degree of desired behaviour. Polynomial interpolation of the subsequent curves then gives the estimated higher order modes, that are  extracted and presented independently of the spurious solutions, as in Fig. 2 (for the second order longitudinal mode). As an example, the second (averaged) longitudinal mode is shown by the dotted black line in Fig.~\ref{fig:FullDisp}, corresponding to that of the first rod in Fig. 2 in the main text. 

For a given frequency of excitation, a transition in the dominant behaviour of the rod motion can be engineered at a selected spatial position by the grading of the array. The introduction of a grading parameter, for example the rod height, can result in an effective bandgap where there is a change in preferential dominance of the rods. It is important to note that this is not a true bandgap resulting from the periodicity or symmetry breaking. Despite this, at the position where the rod's dominant behaviour changes, U-processes dominate and a confined reverse converted beam is obtained. The confinement of the beam of the reversed body wave is dictated by the grading parameter.

\begin{table}[h]
	\begin{tabular}{|l|l|l|l|}
		\hline
		Diameter, $t$ & periodicity, $a$ & initial height, $h_{0}$ & grading, $\Delta h$ \\ \hline
		$0.5 \si{\milli\meter}$& $1.5 \si{\milli\meter}$ & $0.5 \si{\milli\meter}$                        &  $0.05 \si{\milli\meter}$                         \\ \hline
	\end{tabular}
	\caption{Array, and rod, parameters used in simulations and experiments}
	\label{tab:array}
\end{table}

\begin{figure}[hb!]
	\includegraphics[width = 0.48\textwidth]{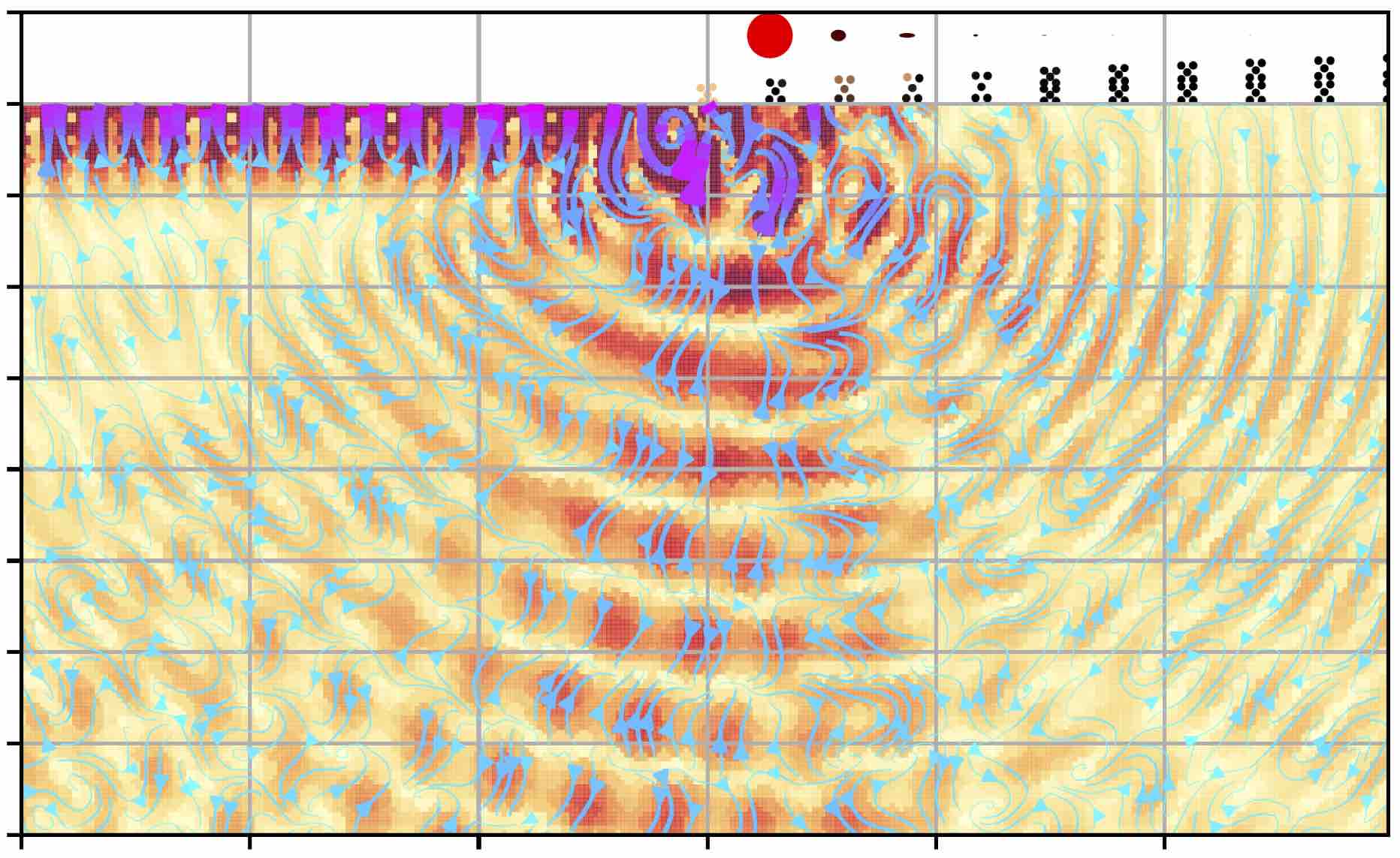}
	\caption{Surface Rayleigh wave to body P-wave conversion directed at $90^\circ$ from the array, described in Table S1, excited at $1.7\si{\mega\hertz}$. The separation of the compressional and shear wave-types is clearly visible from the streamlines.}
	\label{fig:90deg}
\end{figure}

\textbf{\textit{Ungraded arrays:}} As detailed in the main text, Umklapp scattering occurs independently of any grading parameters, so long as the frequencies of excitation result in wavevectors that are outwith the first BZ. This has been shown in electromagnetic systems between regions of abrupt changes in periodicity \cite{Chaplain2019c}; these are not required in the elastic system since surface Rayleigh waves exist at all frequencies independent of any structuring. In light of this we show that Umklapp scattering generates the reversed conversion effect for a single ungraded array of rods, for the same configuration as in Table~S1, but with $\Delta h = 0$. Figure S1(b) shows the isofrequency contours of an incident wave at frequency $1.34 \si{\mega\hertz}$ in purple (obtained from the averaged longitudinal dispersion curves), with the isocircles of the Rayleigh, shear and compressional lines in red, green and black; these are not all defined in the same plane (Rayleigh waves do not propagate in the bulk, and indeed for 1D arrays are points on the horizontal wavenumber axis and S/P waves cannot be localised to the surface), but for clarity we show them superimposed, and extended to circles, to aid the explanation.  Figure~\ref{fig:FullDisp}(c) shows the result of a scattering simulation with no change in resonator parameters: Umklapp scattering is clearly seen, with a mixture of both S and P waves, since there is no effective band gap to confine the conversion into a confined beam of a single wave type. The amplitude of the localised wave is reduced as it transits the array as a result of crystal momentum transfer at every point along the array and this then forms a leaky elastic antenna waveguide. 

\textbf{\textit{Pure P-conversion:}} To exemplify the unprecedented tailored control achieved by simple metawedge structures we demonstrate mode conversion of the surface wave into exclusively a P-wave in the bulk. This is achieved using the graded array presented in the main article, described by the parameters in Table~S1, but now with the frequency increased to $1.7\si{\mega\hertz}$ such that the effective band gap is reached almost immediately along the array. A SPECFEM simulation of this scenario is shown in Fig.~\ref{fig:90deg} whereby, due to the increase in frequency, the resulting reverse converted compressional body wave is produced at nearly $90^{\circ}$ from the array axis. We have therefore achieved  separation of the compressional body waves from any shear waves produced by scattering from the interface, and also been able to tune the angle of conversion. 

\begin{figure}
	\includegraphics[width = 0.48\textwidth]{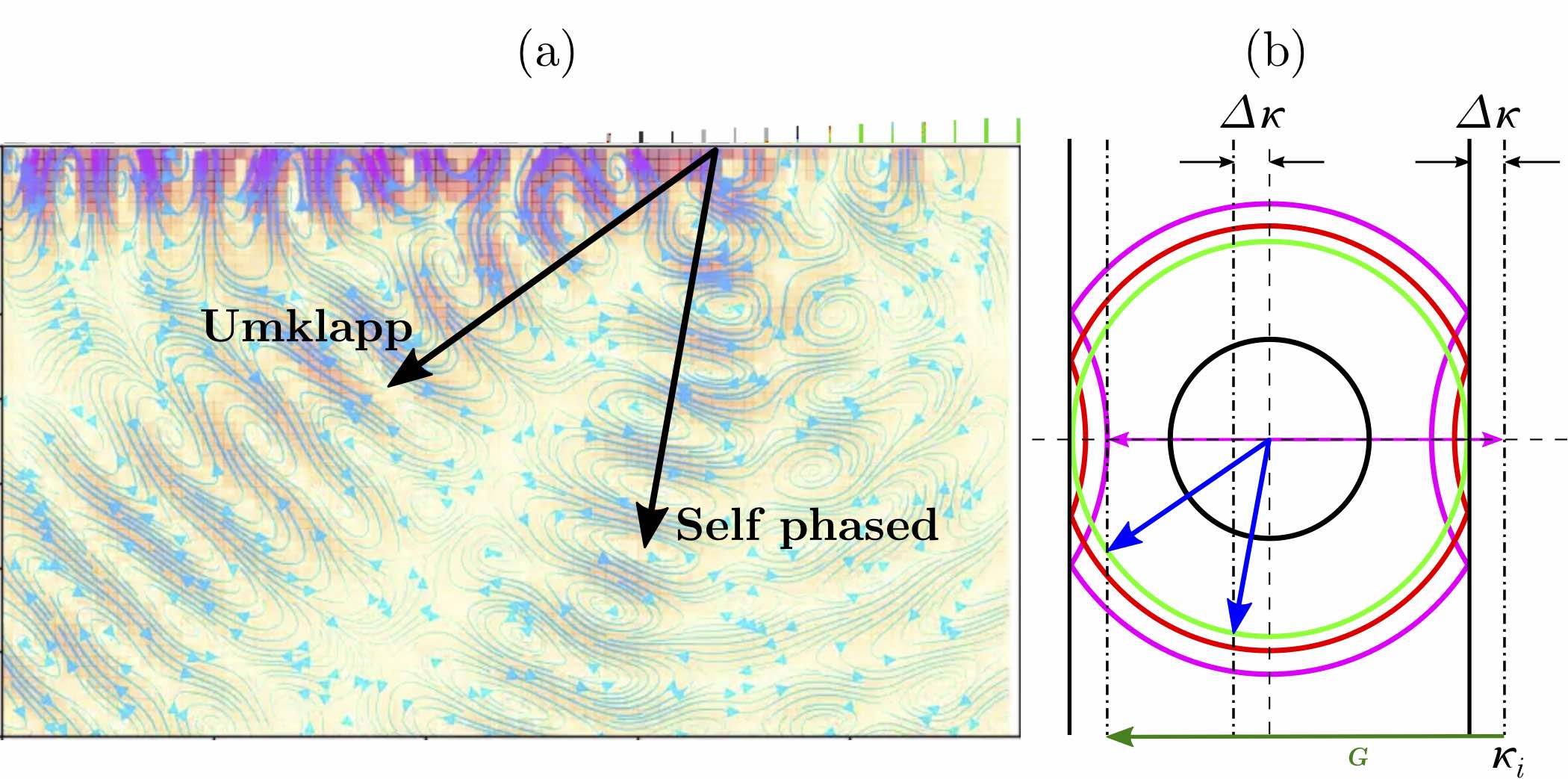}
	\caption{Combining Umklapp and self-phased arrays, resulting in a double S-wave reversed conversion. (a) shows the wave-type of the body wave through the streamlines, whilst (b) combines the Umklapp analysis with that of \cite{Chaplain2019Flat}.
		The incident surface wave has wavenumber $\kappa_i$ and $\Delta\kappa=\vert \kappa_i-\pi/a\vert$ is the difference between it and the wavenumber at the band-edge. 
		The grading is chosen such that the self-phased effect takes place before the effective band gap and cut-off position are met.}
	\label{fig:DoubleS}
\end{figure}

\textbf{\textit{Combining Umklapp and Passive self-phasing:}}
In \cite{Chaplain2019Flat}, for a simpler scalar problem, we developed a passive self-phasing array using graduations of a structured line array. In this section we  combine this with the Umklapp mechanism. Fig.~\ref{fig:DoubleS} shows the quite remarkable splitting of the incident surface wave, where we have an incident array vector of $\kappa_{i}$ (of frequency $1.04\si{\mega\hertz}$), into \textit{two} reversed S-body waves.

The self-phasing effect arises due to the propagation of the surface wave through the graded region until the effective bandgap is reached and the surface wave then slows down and stops.
Once the phase has been altered, such that, in reciprocal space, the frequency and wavenumber are at the band-edge reflection then occurs; upon reflection, the wave is endowed with a phase change of $-\pi/a$, with $a$ being the periodicity.

This reversed surface wave then undergoes similar phase changes as it transverses the array in the opposite direction. Viewing this series of phase changes in isofrequency phase space, as shown in Fig.~\ref{fig:DoubleS}(b), elucidates that the initial $\kappa_{i}$ wave vector is reflected and translated a distance $-\Delta\kappa$ from the origin, and this then co-exists with the U-processes (for which there is a phase shift of $2\pi/a$) resulting in the second reversed shear wave, at a different angle. 

Despite the resultant vector lying within the isocircle of the compressional body wave, see Fig.~\ref{fig:DoubleS}(b), the shear mode is preferentially excited by virtue of the matching between the flexural motion of the rods and the shear waves. The self-phasing effect is intrinsically different to the use of Umklapp scattering; for U-processes we require the transfer of crystal momentum, whereas for passive self-phased arrays we utilise reflection at the band edge. If the grading is designed accordingly, i.e. both rod height and array periodicity altered, the two effects can be separated in space. U-processes take place along the array until the effective band gap is reached (Fig.~S1), and so there is still propagation of energy along the array. It is in this hybrid region where self-phasing can be utilised.

\begin{figure*}[ht!]
	\centering
	\includegraphics[width = 0.95\textwidth]{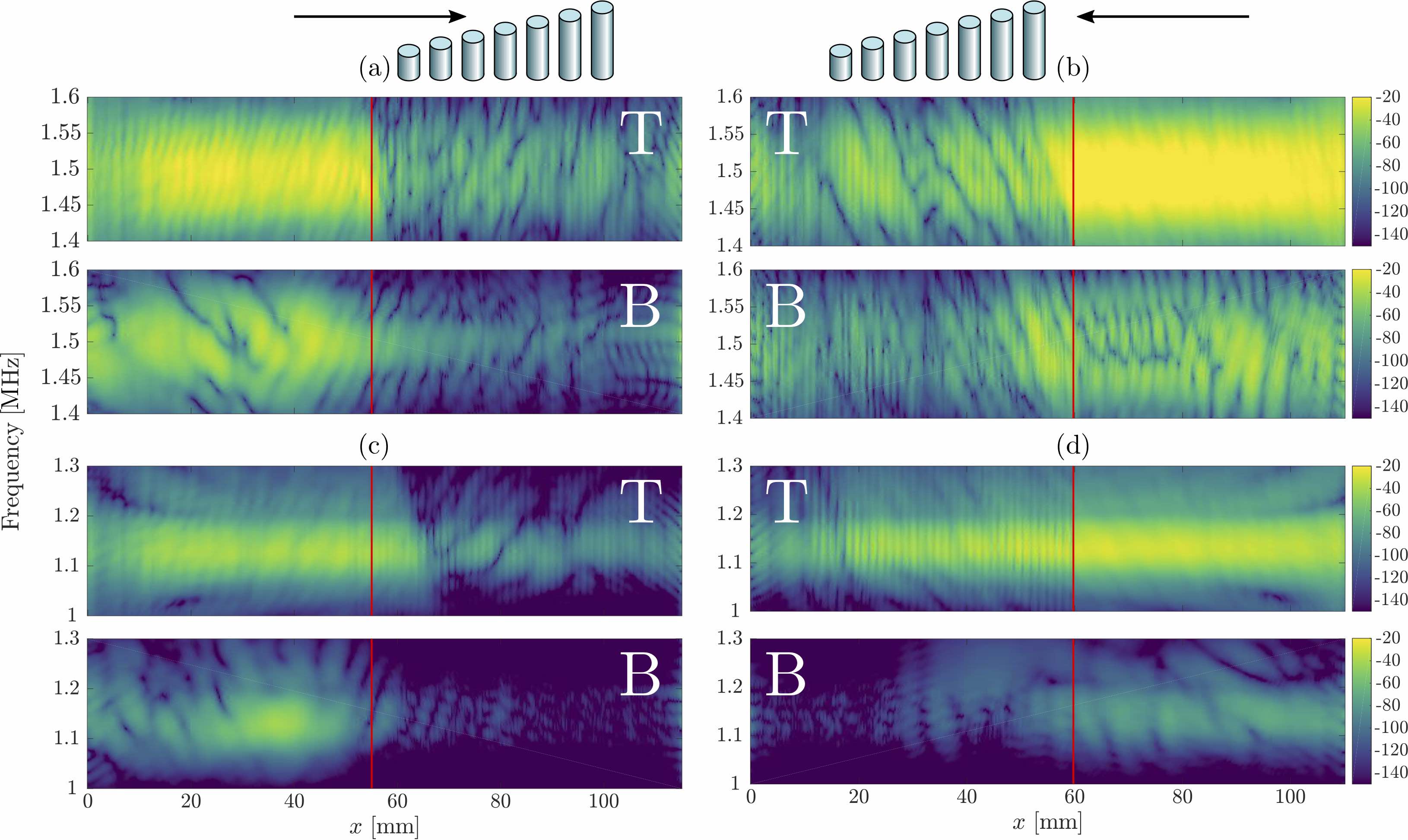}
	\caption{Logarithm of amplitude of Fourier components, extracted from experiments, as function of horizontal position on top and bottom surfaces, marked T,B respectively. Shown above each column is the direction of wave propagation and grading direction of the rods, with array starting from solid red line. (a,b) Show reversed P wave conversion with (c,d) showing reversed S wave conversion, for filtered frequencies between $1.45 - 1.55$MHz and $1.1 - 1.2$MHz respectively. In each case there is a clear reduction in amplitude as the wave on the top surface is reflected and reverse converted by the array. The effect of propagating from short to tall resonator height shows the effective band gap confines the reversed beam compared to exciting from tall to short.}
	\label{fig:FFT_S}
\end{figure*}

\textbf{\textit{Reversing the  grading profile:}} A natural question is how reversing the orientation of the graded array in the main text affects the results, that is, going from tall to short resonators rather than vice-versa. We present here the analysis and experimental results for this case. For these rod heights, there is considerable mixing between the longitudinal and flexural motion of the rods, and there is no distinct separation between longitudinal and flexural rod motions and as a result no clear effective band gap is created. As a result, the wave propagates along the array, experiencing U-processes at every position, since the operational frequencies are still within the second BZ (only the rod heights are changing, not the periodicity). Therefore similar effects to those in the main article occur, except that the reversed Umklapp conversion does not result in a well confined beam; the array behaves as a leaky antenna waveguide, similarly to the ungraded array of Fig.~S1.

Fig.~\ref{fig:FFT_S} allows us to compare and contrast the response as we alter the  orientation of the array relative to the incident surface wave. Here we show comparisons of the logarithms of the amplitudes of the Fourier components measured along the array axis. In the main text we discuss the situation of the surface wave incident upon an array graded from short to tall resonators; Fig.~\ref{fig:FFT_S}(a,c) shows this case for P and S reversed conversion confirming the discussion and interpretation of the main article; there is a sharp conversion from surface wave to body wave (as seen by its impact upon the bottom surface). 

For the opposite case of the wave travelling from tall to short resonator heights, this sharp conversion does not occur and energy is carried to the end of the array, with the conversion effect occuring along the array; an effective bandgap is not reached as there is no clear distinction between longitudinal and flexural motion for the taller rods, shown in Fig.~\ref{fig:3D_top_bottom} (the counterpart of Fig. 1 of the main text). By inspecting the dispersion curves associated with each local rod height the prediction of the angle of reversal is then dictated by the local Umklapp reversed wavevector and the sharp conversion is replaced by a distributed leakage of energy into reversed waves.

\begin{figure*}
	\centering
	\includegraphics[width = 0.95\textwidth]{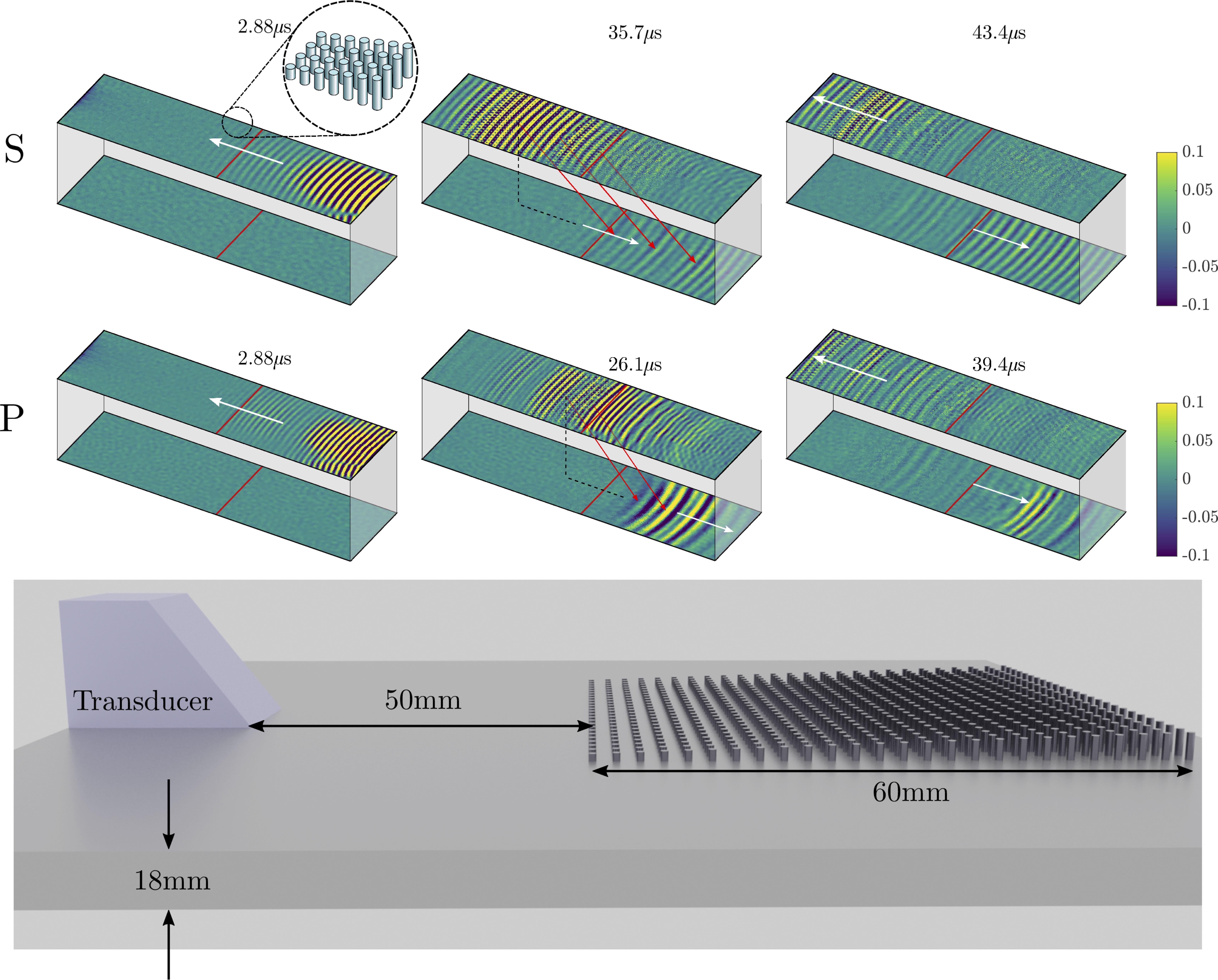}
	\caption{{Experimental results:} snapshots of temporal-spatial filtered scans from tall resonators to short resonators (i.e. a reversed excitation direction compared to that in the main text) along top and bottom surfaces observing S conversion (top row) and P conversion (bottom row), filtered between $1.1-1.2$MHz and $1.45-1.55$MHz respectively. Solid red lines show position where graded array begins, with zoom on the array geometry. Shown too is a schematic of the arrangement between the transducer and the array, with the plate width. \textit{Videos of the experimental results, both filtered and unfiltered are available in the online supplemental material.}}
	\label{fig:3D_top_bottom}
\end{figure*}

\end{document}